\documentstyle[amssymb,preprint,eqsecnum,aps,epsfig,prb]{revtex}
\begin{document}
\draft
\title{Hydrostatic Pressure Effect on the Superconducting Transition Temperature of
MgB$_{2}$}
\author{B. Lorenz, R. L. Meng and C. W. Chu$^{1}$}
\address{Department of Physics and Texas Center for \\
Superconductivity, University of Houston,\\
Houston, Texas 77204-5932\\
$^{1}$also at Lawrence Berkeley National Laboratory, 1 Cyclotron Road,\\
Berkeley, California 94720\\
submitted April 16, 2001}
\date{\today}
\maketitle

\begin{abstract}
The pressure effect on the superconducting transition temperature of MgB$%
_{2} $ has been determined using gas pressure up to 1 GPa. The transition
temperature $T_{c}$ was found to decrease linearly at a constant rate over
the whole pressure range. The recently observed dramatic decrease of $\mid
dT_{c}/dp\mid $ at the 40 K freezing pressure (0.5 GPa) cannot be confirmed.
The pressure coefficient was also found to be independent of the hydrostatic
or nonhydrostatic He environment. The differences in recently reported
values of $dT_{c}/dp$ may be attributed to variations in the sample
conditions, e.g. stoichiometry.
\end{abstract}


The recent discovery \cite{akimitsu} of superconductivity in MgB$_{2}$ at
temperatures as high as 40~K has generated great interest. MgB$_{2}$, which
exhibits an AlB$_{2}$ structure with honeycomb layers of boron atoms,
appears to be electrically and mechanically three-dimensional \cite{kortus}
and its grain boundaries have a far less detrimental effect on
superconducting current transport.\cite{finnemore} The new compound may
provide a way to a higher superconducting transition temperature $T_{c}$ and
an easier avenue for devices. Shortly after the discovery of this exciting
compound a still ongoing discussion was initiated whether the
superconductivity in MgB$_{2}$ is better described by a BCS-like theory\cite
{kortus} or by heavily dressed holes in an almost completely filled
conduction band.\cite{hirsch} The boron isotope effect on $T_{c}$\cite{budko}
and a BCS-like superconducting gap structure\cite{rubio} favor the BCS-type
pairing mechanism. The pressure effect on $T_{c}$ is of special interest
since the dressed hole theory predicted an increase of $T_{c}$ with pressure
as long as there is no charge transfer between the boron and magnesium
planes.\cite{hirsch} First high pressure measurements revealed a negative
pressure coefficient of $dT_{c}/dp\thicksim -1.6\ K/GPa$\cite{lorenz}
indirectly supporting the BCS-mechanism. Subsequent band structure
calculations are in good agreement with the experimental pressure effect and
could explain the decrease of $T_{c}$ within the BCS model by a pressure
induced change of the density of states and the phonon frequency.\cite{loa}
The negative sign and the order of magnitude of the pressure coefficient
were later confirmed but the absolute value $dT_{c}/dp$ varied from $-1.1\
K/GPa$\cite{tomita} to $-2\ K/GPa.$\cite{saito}

Compressibility measurements performed at room temperature show consistently
that the $c/a$ ratio changes very little under pressure (about 1 \% at 10
GPa)\cite{vogt,prassides,jorgensen} indicating nearly isotropic compression.
The same conclusion was drawn from band structure calculations under
pressure.\cite{vogt,loa} Jorgensen et al. recently found that the
compression along the c-axis is 64 \% larger than along the a axis.\cite
{jorgensen} As a result, they proposed that a truly hydrostatic pressure is
indispensable to obtain correct results of $MgB_{2}$. Using a He-gas
pressure system to generate the best hydrostatic environment for $MgB_{2}$,
Tomita et al.\cite{tomita} found that $dT_{c}/dp=-1.11\ K/GPa$ up to $0.5\
GPa$ but drops to almost zero above $0.5\ GPa$. This is in strong contrast
to what previously was observed, namely, $T_{c}$ decreases with pressure
linearly up to $1.8\ GPa$ at a greater rate of $-1.6$ to $-2.0K/GPa$.\cite
{lorenz,saito} Liquid He is known to freeze at about $40\ K$ under $0.5\ GPa$%
. Consequently, Tomita et al. proposed their smaller $\mid dT_{c}/dp\mid $
below $0.5\ GPa$ should be the ''true hydrostatic value'' and the nearly
zero $\mid dT_{c}/dp\mid $ above $0.5\ GPa$ should be a result of the
non-hydrostaticity associated with the freezing of the liquid He pressure
medium. These observations have been cited as supports for the proposed
sensitive role of hydrostaticity in the $T_{c}$-behavior of $MgB_{2}$ under
pressure.\cite{jorgensen,tomita} Unfortunately, the reduction of $\mid
dT_{c}/dp\mid $ above $0.5\ GPa$ due to the proposed shear-stress effects
cannot reconcile with the larger $\mid dT_{c}/dp\mid $ previously observed
in a less-than-ideal hydrostatic environment. It should be noted that,
despite of the greater compression along the $c$-axis than along the $a$%
-axis reported, the overall fractional changes in $c/a$, $c$ and $a$ up to $%
0.6\ GPa$ are very small, $\thicksim 10^{-4}$. Given the small
compressibility of $MgB_{2}$, the drastic $\mid dT_{c}/dp\mid $ change upon
the freezing of liquid He is rather puzzling in the absence of any phase
transition in a quasihydrostatic pressure up to $8\ GPa$, especially in view
of the fact that solid He is the softest material at low temperature. An
experimental artifact due to a failure to deliver pressure to the sample
chamber after freezing of liquid He is therefore suspected.

We have, therefore, carried out high pressure experiments on $MgB_{2}$
samples with different $T_{c}$'s\ up to $1\ GPa$ using helium as pressure
medium. The pressure coefficient of $T_{c}$ is carefully monitored in the
hydrostatic ($p<0.5\ GPa$) region and at higher pressure where the He
freezes above $T_{c}$. $T_{c}$ was found to decrease linearly with pressure
over the whole pressure range at a rate depending on the sample. We conclude
that nonhydrostatic pressure environment has no or only minor effect on the
superconducting transition temperature of $MgB_{2}$. We also conclude that
the observed differences in the value of $dT_{c}/dp$ are due to subtle
differences in sample purity, porosity, or stoichiometry.

For the high pressure measurement we prepared a high quality polycrystalline 
$MgB_{2}$ sample using the standard synthesis. Small Mg chips (99.8 \% pure)
and B powder (99.7 \%) with a ratio of Mg:B = 1.25:2 were sealed inside a Ta
tube in an Ar atmosphere. The magnesium was added in excess of the
stoichiometric amount in order to compensate for any Mg loss during the
synthesis. The sealed Ta ampoule was in turn enclosed in a quartz tube. The
ingredients were heated slowly up to $950\ 
%TCIMACRO{\UNICODE[m]{0xb0}}%
%BeginExpansion
{{}^\circ}%
%EndExpansion
C$ and kept at this temperature for 2 hours, followed by furnace-cooling to
room temperature. The samples so-prepared were dense and x-ray spectra show
a very minor amount of MgO phase. The resistivity and thermoelectric power
of this sample show very sharp transitions to zero at $39.3\ K$ (midpoint of
the superconducting transition) with a width of less than $0.14\ K$. The ac
susceptibility, $\chi _{ac}$, at ambient pressure exhibits an equally sharp
diamagnetic drop at $39.2\ K$ (midpoint) as shown in Fig. 1. For comparison,
we have also re-measured another $MgB_{2}$ sample previously studied under
the same conditions in the helium environment. This sample exhibits a lower $%
T_{c}$ ($<38\ K$) and a broader transition. This sample was previously
investigated in the Fluorinert FC 77 pressure medium.\cite{lorenz}

The superconducting transition was detected by ac susceptibility
measurements. The sample was placed in a transformer in the He gas pressure
cell (UNIPRESS) which was connected to a 1.5 GPa gas compressor (UNIPRESS)
by a beryllium copper capillary (0.3 mm ID). The gas pressure cell and part
of the capillary was inserted into a Model 8CC Variable Temperature Cryostat
(CRYO Industries) for cooling and temperature control. Special care was
taken in cooling at high pressure ($p>0.5\ GPa$) to avoid freezing of helium
in the capillary before it solidifies in the pressure cell. If frozen helium
blocks the capillary first and then solidifies inside the pressure cell a
large drop of pressure\ (about 13 \% at 0.7 GPa) in the cell is usually
observed which may not be recognized if the manometer is located in the room
temperature pressure reservoir. This can easily lead to large errors in the
pressure measurement. Therefore, the pressure cell was cooled very slowly by
controlling the temperature of the cooling He gas to guarantee that the
helium freezes from the bottom of the cell towards the upper end connected
to the gas supply capillary. The cooling process was monitored by two
thermometers mounted to the top and the bottom of the gas pressure cell.
Furthermore, a semiconductor pressure gauge was placed inside the pressure
cell close to the sample position and the pressure was measured $in\ situ$
also in the solid state of the pressure medium. The pressure values used in
Figs. 1 and 2 are measured right at the superconducting transition
temperature.

In the first pressure cycle the cell was loaded to $1\ GPa$ at room
temperature. After cooling and solidification of the helium the pressure
decreased to $0.843\ GPa$ at $T_{c}=38.29\ K$. The cell was heated to above $%
150\ K$ before changing pressure. $\chi _{ac}$ was measured during cooling
and warming through the transition. Fig. 1 shows a set of data taken at
different pressures. The pressure values indicated in the figure refer to $%
p(T_{c})$. The diamagnetic drop of $\chi _{ac}$ shifts in parallel to lower
temperature with increasing pressure. $T_{c}(p)$ was determined as the
midpoint temperature of this drop. As shown in Fig. 2, $T_{c}$ is a linear
function of $p$ over the whole pressure range. The pressure coefficient of $%
-1.07\ K/GPa$ is very close to the value of Tomita et al. \cite{tomita} in
the hydrostatic range ($p<0.5\ GPa$), however, the drastic decrease of $\mid
dT_{c}/dp\mid $observed by them at higher pressures is not detected in our
experiments.

There remains the question if the larger absolute value of $dT_{c}/dp$
observed in the piston-cylinder clamp using quasi hydrostatic pressure media
may be a consequence of pressure induced shear stress. We repeat the He gas
pressure measurement with our $MgB_{2}$ sample that was shown to yield a
pressure coefficient of $-1.6\ K/GPa$ using the Fluorinert FC77 as pressure
medium.\cite{lorenz} Again, $T_{c}$ decreases linearly with $p$ over the
pressure range to $0.84\ GPa$ and no anomaly is detected in passing the
freezing pressure of He. A pressure coefficient $dT_{c}/dp=-1.45\ K/GPa$ is
obtained and is in agreement with our previous data (within the experimental
uncertainty). As mentioned above, this sample shows a lower $T_{c}$ and a
broader transition. We propose that the spread of $dT_{c}/dp$ reported by
different groups\cite{lorenz,saito,tomita} is rather due to subtle
differences in the sample condition, e.g. composition, than to shear stress
in quasi hydrostatic pressure environment.

In conclusion we have shown that the pressure effect on the superconducting
transition temperature of MgB$_{2}$ is linearly negative to the highest
pressure studied and is insensitive to small deviations from truly
hydrostatic pressure conditions. Our results support the view that $MgB_{2}$%
, despite its layered structure, is nearly isotropic with respect to
compression. The variation in the value of $\mid dT_{c}/dp\mid $ by various
groups results from the differences in sample conditions such as composition.

\acknowledgments This work was supported in part by NSF Grant No.
DMR-9804325, the T.~L.~L.~Temple Foundation, the John J.~and Rebecca Moores
Endowment and the State of Texas through the Texas Center for
Superconductivity at the University of Houston; and at Lawrence Berkeley
Laboratory by the Director, Office of Energy Research, Office of Basic
Energy Sciences, Division of Material Sciences of the U.~S.~Department of
Energy under Contract No. DE-AC03-76SF00098.

\begin{figure}[tbp]
\caption{$\protect\chi_{ac}$ {\it vs.} $T$ at various pressures.\newline
open symbols: data taken in liquid He; closed symbols: data taken in solid
He }
\label{XT}
\end{figure}

\begin{figure}[tbp]
\caption{$T_{c}$ as function of pressure.}
\label{TP}
\end{figure}

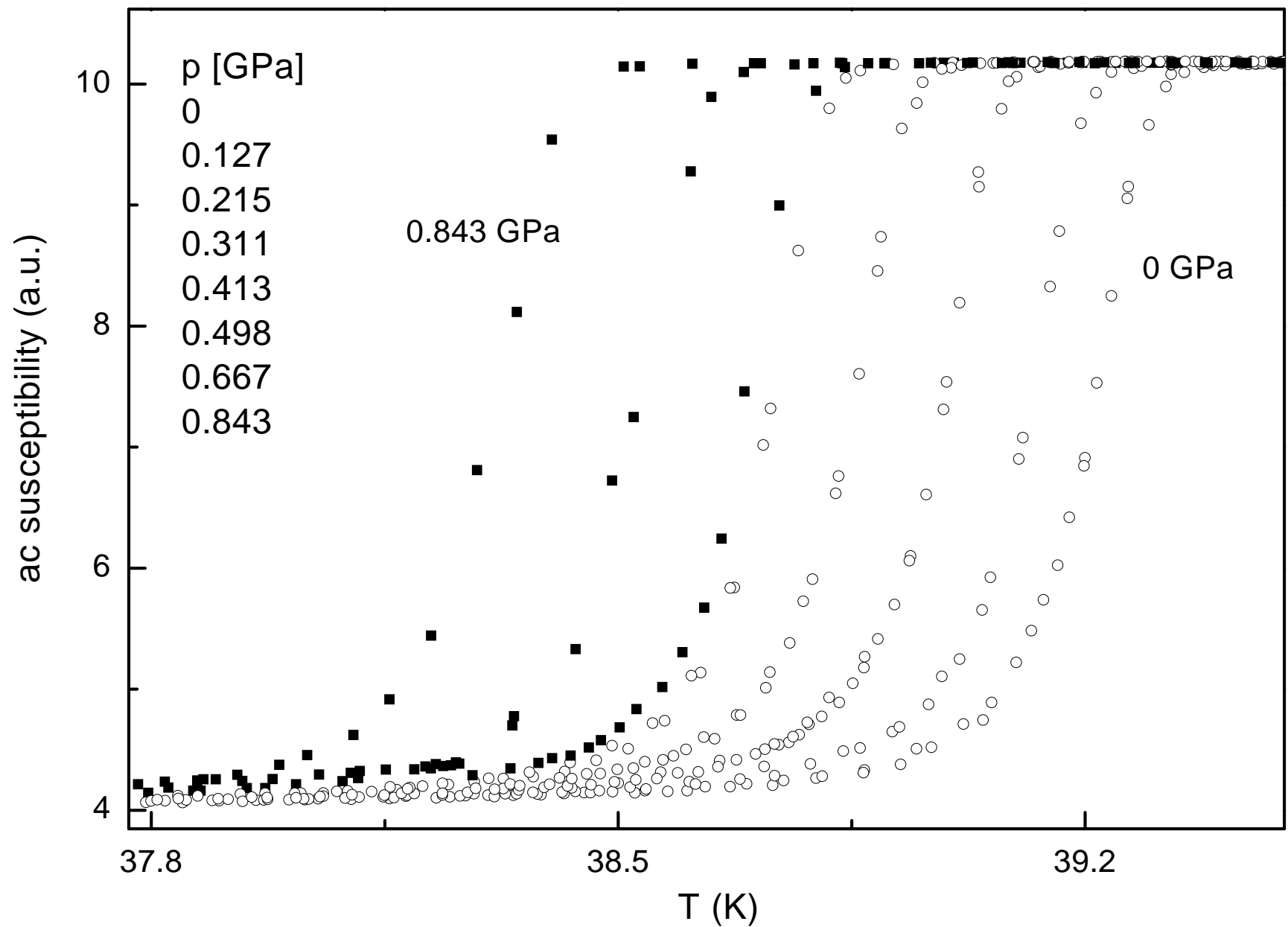
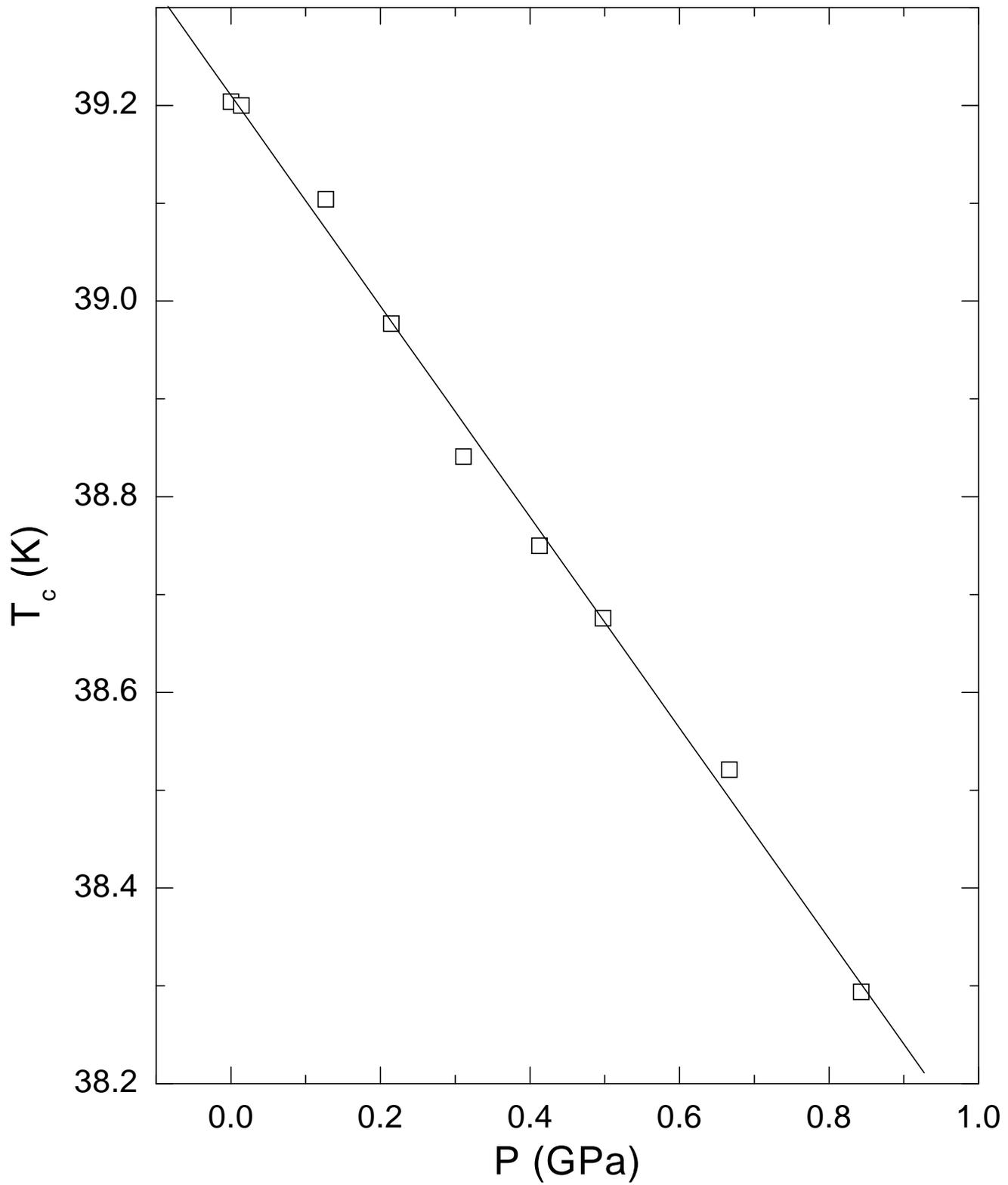
\end{document}